\documentclass[traditabstract,letter]{aa}
\usepackage{txfonts}
\usepackage{graphicx}
\usepackage{natbib}
\bibpunct{(}{)}{;}{a}{}{,} % to follow the A&A style

\newcommand{\beq}{\begin{equation}}
\newcommand{\eeq}{\end{equation}}
\newcommand{\ba}{\begin{eqnarray}}
\newcommand{\ea}{\end{eqnarray}}
\def\HI{\hbox{H~$\scriptstyle\rm I$}~}
\def\HII{\hbox{H~$\scriptstyle\rm II$}~}

\def\nHII{{\rm HII}}

\def\HeII{\hbox{He~$\scriptstyle\rm II$}}

\def\spose#1{\hbox to 0pt{#1\hss}}
\newcommand{\lta}{\mathrel{\spose{\lower 3pt\hbox{$\mathchar"218$}}
      \raise 2.0pt\hbox{$\mathchar"13C$}}}
\newcommand{\gta}{\mathrel{\spose{\lower 3pt\hbox{$\mathchar"218$}}
      \raise 2.0pt\hbox{$\mathchar"13E$}}}
\def\lsim{\mathrel{\rlap{\lower 3pt\hbox{$\sim$}}\raise 2.0pt\hbox{$<$}}}
\def\gsim{\mathrel{\rlap{\lower 3pt\hbox{$\sim$}} \raise 2.0pt\hbox{$>$}}}

\begin{document}

\title{High redshift AGNs and HI reionisation: limits from the unresolved X-ray background}

\author{Francesco Haardt\inst{1,2} \and
Ruben Salvaterra\inst{3}}

\institute{DiSAT, Universit\`a dell'Insubria, Via Valleggio 11, I-22100 Como, Italy \and
INFN, Sezione di Milano-Bicocca, Piazza delle Scienze 3, I-20123 Milano, Italy \and
INAF, IASF Milano, via E. Bassini 15, I-20133 Milano, Italy}

\date{Received/Accepted}

\abstract{The rapidly declining population of bright quasars at $z\gsim 3$ appears to make an increasingly small contribution to the ionising background at the \HI Lyman limit. It is then generally though that massive stars in (pre-)galactic systems may 
provide the additional ionising flux needed to complete \HI reionisation by $z\gsim 6$. A galaxy dominated background,  however, may require that the escape fraction of Lyman continuum radiation from high redshift galaxies is as high as 10\%, a value somewhat at odds with (admittedly scarce) observational constraints. High escape fractions from dwarf galaxies have been advocated, or, alternatively, a so-far undetected (or barely detected) population of unobscured, high-redshift faint AGNs. Here we question the latter hypothesis, 
and show that such sources, to be consistent with the measured level of the unresolved X-ray background at $z=0$, can provide a fraction of the \HII filling factor not larger than 13\% by $z\simeq 6$. The fraction rises to 
$\lsim 27\%$ in the somewhat extreme case of a constant comoving redshift evolution of the AGN emissivity. This still calls for a mean escape fraction  of ionising photons from high-$z$ galaxies $\gsim 10\%$.} 


\authorrunning{Haardt \& Salvaterra}
\titlerunning{XRB limits to reionisation}

\maketitle

\section{Introduction}

The reionisation of the all-pervading intergalactic medium (IGM) is a landmark event in the history of the Universe. 
Studies of the so-called Gunn-Peterson absorption in the spectra of distant quasars show that hydrogen was already highly ionised out to redshift $z\sim 6$ \citep[e.g., ][]{songaila04,fan06a}, while CMB polarisation data constrain the redshift of a sudden reionisation event to be significantly higher, $z\sim 10$ \citep{jarosik11,hinshaw13}.

%Despite much recent progress, a coherent description of the thermal state and ionisation degree of the IGM, the repository of most of the baryons across the history of the Universe, remains elusive. 
Most of our understanding of IGM physics, and its implication for galaxy formation and metal enrichment, depends critically on the properties of the cosmic ionising background. 
%Existing photoionisation models suggest that, at $z\simeq 3$, the hydrogen neutral fraction at cosmic mean density is only
% $\sim 10^{-5.5}$ \citep{kollmeier03}, and 
 While it is generally thought that the gas is kept ionised by the integrated UV emission from 
 active galactic nuclei (AGNs) and star-forming galaxies \citep{miralda90,HM96}, the relative contributions 
 of these sources as a function of  cosmic time are poorly known. 

% Because of the high ionisation threshold and small 
% photoionisation cross-section of HeII, and of the rapid recombination rate of HeIII, the double ionisation of helium is expected to 
% be completed by hard UV-emitting quasars around the peak of their activity, at $z\simeq 2.5$ \citep{madau94b,sokasian02,mcquinn09}, much later than the reionisation of \HI and \HeI.  
 At $z\gsim 3$, the declining population of  bright quasars appears to make an increasingly small contribution to the ionising radiation background at the \HI Lyman limit. It was then suggested that massive stars in galactic systems may provide the additional ionising flux needed at early times \citep[e.g.][]
 %{MHR99,gilmore09,robertson10}. 
{MHR99,gnedin00,wyithe03,meiksin05,trac07,faucher08a,gilmore09,robertson10}. 
 However, leaking Lyman continuum radiation from bright galaxies seem to be 
 modest \citep[see, e.g.,][]{vanzella10}, and it has been therefore argued that dwarf galaxies 
 (with virial mass below $\sim 10^9$ M$_\odot$) 
 may produce the dominant contribution to the \HI ionising UV background \citep[e.g.,][]{robertson12}. 

%A further indication that galaxies responsible of the reionisation of the IGM are of the sub-$M_\star$ flavour comes from the analysis of 
%the required photon budget. A simple estimate of the minimum amount of high energy photons needed to complete reionisation 
%by $z\sim 7$ (i.e., one photon per baryon per recombination time), once converted into a minimum dark 
%matter halo mass, would result in the requirement of vigorous star formation in dark matter haloes down to $10^8$ M$_\odot$ \citep{boylan14}. Similarly, it has been shown that a phenomenological model of the ionising background able to 
%sustain reionisation by $z\sim 6-7$ requires that a steep Schechter-like luminosity function is integrated down to $0.01 L_\star$, with an escape fraction of \HI ionising radiation as large as $\sim 10\%$ \citep{HM12}. The more recent determination of the galaxy UV luminosity function (LF) at high redshifts confirms such estimate \citep{finkelstein14}. 

Alternative to invoking a major contribution to reionisation from dwarf(ish) galaxies  is the possibility that the AGN emissivity 
at $z \gsim 4$ is indeed much larger than generally thought. Indications along such line have been reported by several 
groups \citep{glikman11,civano11,fiore12}, though it is fair to say that results seem not so univocal \citep[see, e.g.,][]{masters12}. Very recently, \citet{giallongo14} found 22 AGN candidates at $z\gsim 4$ in the Candel/GOOD-S/Chandra Deep Field South field, suggestive of a prominent contribution of AGNs to the ionising background in the range $4\lsim z \lsim 6.5$.
The resulting \HI photoionisation rate is indeed consistent with various estimates at the same redshifts, based on both the flux-decrement and proximity effect techniques \citep{becker07,calverley11}. 

The high redshift population of AGNs should leave an imprint in the observed cosmic X-ray background (XRB). {\it Chandra} deep observations resolved the XRB into discrete sources at a level of $80-90\%$ over  the entire bandwidth ($0.5-2$ keV), 
with only a fraction $\sim 1\%$ of the signal arising from sources located at $z\gsim 4$ \citep{xue11}.  
\citet{moretti12} exploited the very low instrumental noise of the {\it Swift} XRT to measure the still unresolved XRB 
spectrum at the highest accuracy. In \citet{salvaterra12} we used such measures to place upper limits on the cosmic accretion 
history of massive black holes. An obvious caveat to our conclusions is the possible existence of a large population of severely obscured ($\log N_H\gsim 25$)accreting black holes, hence not glowing in the X-rays. Unless advocating a very peculiar UV-to-X-ray spectral energy distribution, such caveat would not apply if the unresolved AGN population does contribute significantly to the ionisation background. The assessment of the contribution to the XRB of such UV-emitting AGNs is precisely the goal of this {\it Letter}. Specifically, we will translate the \citet{moretti12} upper limits to the unresolved XRB into upper limits on the possible contribution of high-redshift, unobscured faint AGNs to \HI reionisation. A similar analysis was proposed by \citet{dijkstra04}, \citet{salvaterra05,salvaterra07}, and \citet{mcquinn12}, with conflicting results. Here we use the most updated limits on the XRB  and adopt a $(h,\Omega_m,\Omega_\Lambda)=(0.7,0.3,0.7)$ cosmology.

\section{Methodology}
%The \citet{moretti12} measurements impose strict limits to the number of ionising photons emitted by an hypothetical population of high redshift, unobscured AGNs thought to be responsible of the reionisation of the hydrogen component of the IGM. A simple estimate of the contribution to the XRB of such unknown population can be obtained as follows. 

Assuming an AGN comoving X-ray specific emissivity $\propto E^{-\alpha_x}(1+z)^{-\gamma}$, 
the XRB at observed energy $E_0$ due to sources located at redshift $z \geq z_x$ is:
%%%%%%%%
\begin{equation}
J_{E_0}(\geq z_x) =\frac{c}{4\pi}\int_{z_x}^\infty {dz\,\left| \frac{dt}{dz}\right| \, \epsilon_{2{\rm keV}}\left( \frac{E}{E_{2{\rm keV}}}\right)^{-\alpha_x}\, 
(1+z)^{-\gamma}},
\label{eq:Jx}
\end{equation}
%%%%%%%%
where $E=E_0(1+z)$. We now  relate the specific emissivity at 2 keV, $\epsilon_{2{\rm keV}}$, to that at 912\AA, i,.e., $\epsilon_{2{\rm keV}}=K\epsilon_{912\AA}$, where the ``{\it K}-correction" normalisation reads
%%%%%%%%
\begin{equation}
K= \left( \frac{1300\AA}{912\AA}\right)^{\alpha_{\rm fuv}}\,
\left( \frac{2500\AA}{1300\AA}\right)^{\alpha_{\rm uv}}\,
\left( \frac{E_{2{\rm keV}}}{E_{2500\AA}}\right)^{-\alpha_{\rm ox}}. 
\label{eq:Kcor} 
\end{equation}
%%%%%%%%
Here $\alpha_{\rm ox}$ is the optical-to-X-rays spectral index, defined from the specific emissivity at 2 keV and at 2500 \AA, $\alpha_{\rm ox}\equiv -0.384 \log(\epsilon_{2 {\rm keV}}/\epsilon_{2500 \AA})$. In writing eq.~\ref{eq:Kcor} we followed  
the piece-wise UV AGN spectral energy distribution as described in \citet{HM12}. 
Now the r.h.s. integral in eq.~\ref{eq:Jx} is easily solved, and the obtained $J_{E_0}(\geq z_x)$ constrained to be not larger than the observational upper limit $J_{E_0}^{\rm obs}$. This in turn gives the maximum value of $\epsilon_{912\AA}$ consistent with the limits on the unresolved XRB:
%%%%%%%%
\begin{equation}
\epsilon_{912\AA}(z) \leq \frac{4\pi H_0 \Omega_m^{1/2}}{c\,K\,R_{II}}\,\eta
(1+z_x)^\eta \,(1+z)^{-\gamma}\,J_{E_0}^{\rm obs}, 
\label{eq:eps912}
\end{equation}
%%%%%%%%%
where we neglected the energy density of the cosmological constant (we are interested in the redshift regime $z\gg z_{m\Lambda}\simeq 0.33$). We set $\eta \equiv (\gamma+\alpha_x+3/2)$ and the term $R_{II}\geq 1$ is meant to account for the contribution of obscured AGNs at $z\geq z_X$ to the XRB observed at energy $E_0$. 

It is now straightforward to translate the above limit into a limit on reionisation. The volume filling factor of \HII regions $Q_\nHII$ is the solution of the following differential equation \citep[see][]{MHR99}:
%%%%%%%%
\begin{equation}
\frac{dQ_\nHII}{dt}=\frac{\dot n}{n_0}-\frac{Q_\nHII}{t_{\rm rec}},
\label{eq:Qvsz}
\end{equation}
%%%%%%%%
where $n_0$ is the cosmic hydrogen mean density, and $\dot n(z)=\epsilon_{912\AA}(z)/(h_p\alpha_{\rm fuv})$ 
the photon emission rate ($h_p$ is the Planck constant, and the FUV 
emissivity is $\propto \nu^{-\alpha_{\rm fuv}}$). The \HII recombination time $t_{\rm rec}$ is computed as in  
\citet{HM12}.

%The volume filling factor of \HII regions can be approximatively computed as \citep{barkana01}
%%%%%%%%%
%\begin{equation}
%Q_{\rm HII}(z)=\int_{z}^{\infty}{dz' \,  \left| {\frac{dt}{dz'}}\right| \frac{\dot n(z')}{n_0}F(z,z')},
%\label{eq:qvsz}
% \end{equation}
%%%%%%%%%
%where $n_0$ is the hydrogen mean density, and the function $F(z,z')$ takes into account the effects of \HII recombinations. 
%%As an example, by taking $F(z,z')=1$, eq.~\ref{eq:qvsz} gives the number of ionising photons per hydrogen atom emitted up to redshift $z$. 
%The number of ionising photons emitted per unit volume per unit time $\dot n(z')$ is
%%%%%%%%%
%\begin{equation}
%\dot n(z)=\int_{\nu_{912\AA}}^{\infty} {d\nu \, \frac{\epsilon_E(z)}{h\nu}=\frac{\epsilon_{912\AA}}{h\alpha_{\rm fuv}}(1+z)^{-\gamma}}.
%\label{eq:ndot}
%\end{equation}
%%%%%%%%%
%Substituting into eq.~\ref{eq:qvsz} gives a relation between $Q_{\rm HII}$ and $\epsilon_{912\AA}$:
%%%%%%%%%
%\begin{equation}
%\epsilon_{912\AA}(z)=h\alpha_{\rm fuv}\,\Omega_m^{1/2}\,H_0\,n_0\,I(z)\,Q_{\rm HII}(z),
%\label{eq:epsL}
%\end{equation}
%where 
%%%%%%%%%
%\begin{equation}
%I(z)\equiv \left[ \int_{z}^{\infty}{dz'\, (1+z')^{-\gamma-5/2}F(z,z')} \right]^{-1}.
%\end{equation}
%%%%%%%%%
%From eq.~\ref{eq:eps912} we finally derive the upper limit on the \HII filling factor at redshift $z$:
%%%%%%%%%
%\begin{equation}
%Q_{\rm HII}(z) \leq \frac{4\pi}{ch\alpha_{\rm fuv}\, n_0\, I(z)\, K\, R_{II}} 
%\eta (1+z_x)^{\eta}
%\left(\frac{E_0}{E_{2{\rm keV}}}\right)^{\alpha_x}\,J_{E_0}^{\rm obs}.
%\label{eq:Qvsz}
%\end{equation}
%%%%%%%%%

\section{Observational Parameters}
The upper limits given by eq.~\ref{eq:eps912} and eq.~\ref{eq:Qvsz} depend upon a number  of parameters which need to be observationally constrained. In this section we discuss the choice we make for each of them.

%%%%%%%%%%%%%%%%%%%%%%%%%%%
%\begin{itemize}

%\item 
{\it Unresolved XRB.} 
The unresolved XRB shows a very hard spectrum, suggesting that most (if not all) of the flux comes from low-$z$ obscured AGNs. Indeed, \citet{moretti12}, by adopting the XRB synthesis model of \citet{gilli07}, derived a stringent limit $J_{E_0}^{\rm obs}\leq 1.9 \times 10^{-27}$ erg cm$^{-2}$ s$^{-1}$ Hz$^{-1}$ sr$^{-1}$ at $E_0=1.5$ keV once accounting for absorbed AGNs at $z\lsim 5$ whose fluxes lie below the {\it Chandra} limit. Yet, the synthesis model falls short at $E_0\gsim 3$ keV, suggesting the possible existence of a population of Compton thick AGNs at intermediate redshifts. 

%\item
{\it UV and X-ray spectral indices.} As already stated, we use the very same parametrisation of \citet{HM12}, adopting for the UV slope $\alpha_{\rm uv}=0.44$ ($\lambda>1300$ \AA, \citet{vandenberk01}), and $\alpha_{\rm fuv}=1.57$ in the FUV range ($\lambda<1300$ \AA, \citet{telfer02}). Concerning the X-ray spectrum, we notice that most of the contribution to the XRB at $E_0=1.5$ keV is expected from source located just above $z\simeq 5$, i.e., 
from photons emitted at rest-frame energy $\simeq 10$ keV. In this energy range unobscured AGNs exhibit a power-law spectrum with index $\alpha_x\simeq 0.8$, as a combination of the intrinsic continuum and of the Compton reflection 
bump (see, e.g., \citet{ueda14}). In our analysis we then take $\alpha_x=0.8$. From eq.~\ref{eq:Kcor} it is apparent how the exact values of the UV and X-ray spectral indices will affect our conclusions only marginally. 

%\item
{\it Optical-to-X-ray spectral index.} The value of $\alpha_{\rm ox}$ has a major impact on our estimate of $Q_{\rm HII}$ since $(E_{2{\rm keV}}/E_{2500\AA})\simeq 403$. The study of the correlation between X-ray and UV luminosities has been the subject of many works on both optically selected and X-ray selected AGNs. Among others, \citet{steffen06} found a significant correlation between $\alpha_{\rm ox}$ and the monochromatic luminosities at 2500 \AA~ (i.e., $L_{2500 \AA}\propto L_{2 {\rm keV}}^\beta$, with $\beta>1$) in a sample of 333 optically selected AGNs. No significant correlation of $\alpha_{\rm ox}$ with redshift was reported. \citet{lusso10} analysed a sample of 545 X-ray selected Type I AGNs from the XMM-COSMOS survey, finding again a correlation between $\alpha_{\rm ox}$ and $L_{2500 \AA}$. The mean value of $\alpha_{\rm ox}$ for the full sample was $1.37$ with a dispersion around the mean of 0.18. \citet{marchese12} also reported a highly significant correlation between $\alpha_{\rm ox}$ and the UV luminosity in a sample of 195 X-ray selected Type I bright AGNs, basically confirming the \citet{lusso10} results. In our investigation the supposedly unaccounted population of AGNs responsible of \HI reionisation must necessarily resides 
in the very faint-end of the UV luminosity function. According to the literature cited above, this would imply a value 
of $\alpha_{\rm ox}$ on the lowest side of the distribution, though it must be considered that the redshift range of interest here is basically un-explored at X-ray wavelengths. Given that, we adopt a fiducial value $\alpha_{\rm ox}=1.35$. We are confident that, if the observed correlation between 
$L_{2500 \AA}$ and $\alpha_{\rm ox}$ holds at very high redshifts, such choice is conservative, hence strengthening our conclusions.   

%\item
{\it Redshift evolution.} The evolution of the AGN space density at high redshifts has been subject of several revisions in the last decade, mainly because of the dearth of data at $z\gsim 4$. As an example, \citet{ueda03} adopted, for very luminous sources, an evolution factor $\propto (1+z)^{-\gamma}$ with $\gamma=1.5$ above $z = 1.9$, while for fainter AGNs the turn over occurs at increasingly lower redshift. \citet{silverman08} found a much sharper decline, $\gamma=3.27$, similar to that derived in studies of optically selected QSOs. Recently, \citet{hiroi12} claimed an even stronger decline at $z \gsim 3$, $\gamma=6.2$, a value adopted by \citet{ueda14} in the most recent and updated study of the hard X-ray LF. The situation is somewhat more confusing in the optical-UV band. While different groups agreed on the faint-end slope of the LF, $\simeq 1.7$, they 
sorted out quite different absolute space densities. Specifically, \citet{glikman11} claim roughly a factor four more sources at $z\gsim 3$ compared to \citet{ikeda11}.  More recently, \citet{masters12} found  a 
decrease by a factor of four in the number density of faint QSOs in COSMOS between $z\sim 3.2$ and $z\sim 4$, supporting the results of \citet{ikeda11}. Overall, the results from \citet{masters12} suggest a similar 
evolution of the UV and X-ray LFs at $z\gsim 3$. However, a large normalisation of the UV LF, basically consistent with \citet{glikman11} though at higher redshifts 
($z\simeq 5-6$) and fainter UV magnitudes, was recently claimed \citep{giallongo14}. 
Given these uncertainties, we will assume   
a redshift evolution of the emissivity as sharp as $\gamma=6.2$ in both the X-ray and UV 
bands, but we will also show results for the somewhat extreme case of a constant comoving emissivity ($\gamma=0$). 

%\item
{\it Obscured sources.} The parameter $R_{II}$ in eq.~\ref{eq:eps912} is meant to account for the contribution to the XRB given by sources obscured in the optical band, thus not contributing to the ionising background. Such contribution could be relevant since photons observed at $E_0=1.5$ keV are emitted, by $z\gsim 5$ AGNs, at rest frame energies $\gsim 10$ keV, where the emission of absorbed AGNs is anyway relevant. We implemented the X-ray LF of \citet{ueda14}, and found that $J_{E_0}(\geq z_x)$ (with $E_0=1.5$ keV and $z_x=5$) is almost evenly divided between objects with $\log N_H<22$ and objects with $\log N_H>22$, which would give $R_{II}\simeq 2$. However, it is not simple to determine above which X-ray determined equivalent hydrogen column density $N_H$ sources are severely obscured in the optical-UV band. In their study, \citet{masters12} found that  
$\sim75\%$ of X-ray bright AGNs at $z\sim 3-4$ are indeed optically obscured. Taken at face value, this would imply  $R_{II}\simeq 4$. Finally it is worth noticing that at lower redshifts ($z\lsim 3$) the incidence of obscured AGNs 
is strongly anti-correlated with X-ray luminosity \citep{merloni14}. Provided that the trend is similar at 
earlier epochs, this very fact points toward a high $R_{II}$ correction factor. In our fiducial model we then assume $R_{II}=4$.

%\item
{\it Lower redshift of unresolved XRB.} In our analysis, the limiting redshift $z_x$ plays an important role, as it sets 
the minimum redshift of the unaccounted AGN population we are testing. Clearly, such population must give a contribution to the XRB not exceeding the measured unresolved fraction. Specifically, the upper limits to the unresolved XRB given by \citet{moretti12} were obtained subtracting to the total XRB all sources listed in the 4Ms-Chandra 
catalog \citep{xue11}, which basically contains no AGNs with $z\gsim 5$. As and example, among the 6 AGN candidates at $z\gsim 5$ found by \citet{giallongo14}, only 2 are in the \citet{xue11} catalog. 
Should the unresolved XRB arise from high-$z$ AGNs, it must necessary come from $z\gsim 5$ sources unless 
they own a very peculiar redshift distribution. Given that, we assume 
$z_x=5$ as the lower limiting redshift in our study. 

%\end{itemize}
%%%%%%%%%%%%%%%%%%%%%%%%%%%

To summarise, our benchmark model adopts: $\alpha_{\rm fuv}=1.57$, $\alpha_{\rm uv}=0.44$, $\alpha_x=0.8$, $\alpha_{\rm ox}=1.35$, $\gamma=6.2$, $R_{II}=4$, $z_x=5$, and the XRB limit 
$J_{1.5{\rm keV}}^{\rm obs}\leq 1.9 \times 10^{-27}$ erg cm$^{-2}$ s$^{-1}$ Hz$^{-1}$ sr$^{-1}$.

\section{Results}
%Our study is aimed at placing upper limits on the possible contribution of AGNs to the cosmic ionising background exploring existing upper limits on the unresolved fraction of the XRB. Eq.~\ref{eq:eps912} and eq.~\ref{eq:Qvsz} derived in Section 2 translate such XRB limits into limits on the \HII volume filling factor and on the AGN emissivity at the Lyman edge. 
%%%%%%%%%%%%%%%%%%%%%%%
\begin{figure}
\resizebox{\hsize}{!}{\includegraphics{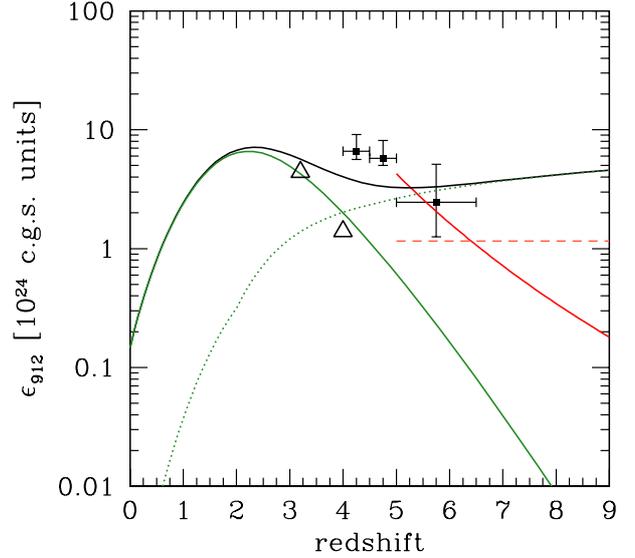}}
\caption{The maximum emissivity at the \HI Lyman limit vs. redshift. The red solid curve at $z\geq 5$  
is our benchmark case, consistent with the XRB limit $J_{1.5{\rm keV}}^{\rm obs}\leq 1.9 \times 10^{-27}$ erg cm$^{-2}$ s$^{-1}$ Hz$^{-1}$ sr$^{-1}$. Such value accounts for sources at 
$z\lsim 5$ below the Chandra flux limit, estimated from the \citet{gilli07} XRB synthesis model. The dashed line shows  the extreme case of a constant comoving redshift evolution of the AGN emissivity.  The black line is the emissivity employed in the UV background model of  \citet{HM12}, with galaxies (dotted green line) and AGNs (solid green line) shown separately. The black data points are the AGN emissivity as estimated by \citet{giallongo14}, while open triangles are data from \citet{masters12}.}
\label{fig1}
\end{figure}
%%%%%%%%%%%%%%%%%%%%%%%
%%%%%%%%%%%%%%%%%%%%%%%
\begin{figure}
\resizebox{\hsize}{!}{\includegraphics{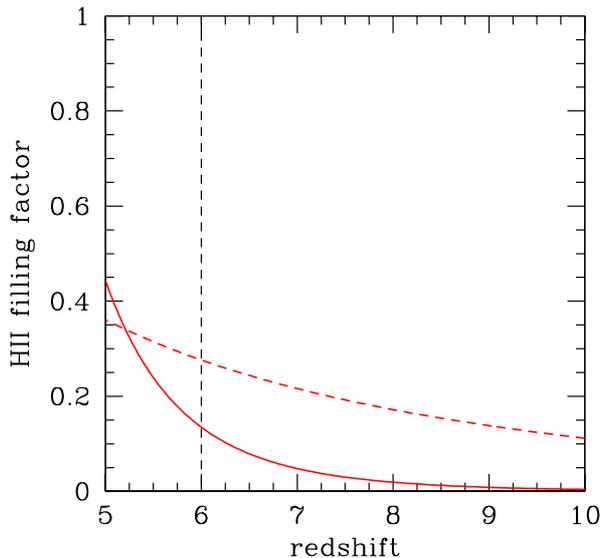}}
\caption{Maximum volume filling factor of \HII vs. redshift. Curves as in Fig.~\ref{fig1}. The dashed vertical line marks a fiducial reionisation redshift, $z=6$. In our benchmark case the contribution to \HII reionisation from AGNs is below 13\%.}
\label{fig2}
\end{figure}
%%%%%%%%%%%%%%%%%%%%%%%
Fig.~\ref{fig1} shows $\epsilon_{912\AA}$ given by eq.~\ref{eq:eps912}. The emissivity at the Lyman limit is a most interesting quantity, as it can be compared to diverse observational estimates existing in literature. Our benchmark case is shown as the red  solid line starting from $z=5$. Our 
limit is  compared to the recent values of \citet{giallongo14} and to the values reported by \citet{masters12} (shown as black data points and open triangles, respectively). An assessment of the reasons behind the discrepancy between these different results is beyond the scope of this letter. Still, taking the \citet{giallongo14} ionising emissivity at face value, we must conclude that the associated AGNs basically saturate the observed XRB, as apparent from Fig.~\ref{fig1}. 

We also compare our estimate of $\epsilon_{912\AA}$ with the ``minimum reionisation model" of \citet{HM12}. 
In Fig.~\ref{fig1} the overall Lyman limit emissivity of \citet{HM12} is shown as a black solid line, along with the separate contribution of AGNs (green solid line) and 
star forming galaxies (green dotted line). The AGN emissivity closely fits the results by \citet{hopkins07}, while for galaxies \citet{HM12} assumed that the fraction of Lyman continuum photons leaking into the IGM is a strong increasing function of redshift. Such model reionises \HI by $z\simeq 6.7$ and \HeII~by $z\simeq 2.8$. Though 
our estimate benchmark emissivity at $z\simeq 5-6$ is similar to the \citet{HM12} model, the approximatively constant comoving  behaviour of the 
latter compared to the steep decline we adopt here leads to the different outcome in terms of reionisation (see next). For such reason we tested the somewhat extreme case of a constant comoving emissivity, shown as a dashed line in Fig.~\ref{fig1}. The resulting maximum emissivity is lower than the \citet{HM12} reionisation model by a factor of $\simeq 4$ at high-$z$. 

In Fig.~\ref{fig2} we show the resulting volume fraction occupied by \HII regions (eq.~\ref{eq:Qvsz}). Our benchmark case allows  for a fraction $\lsim 13\%$ of the IGM to be ionised by AGNs by $z\gsim 6$, showing that AGNs alone can not reionise the Universe. We checked that this conclusion holds in spite of the uncertainties of the parameters. $Q_\nHII=1$ at $z=6$ can be reached only for $\alpha_{\rm ox}\gsim 1.7$, or $\alpha_{\rm ox}\gsim 1.5$ assuming $R_{II}\simeq 1$ (i.e., no obscured sources). Such figures  seem to be unlikely when compared to available data. Finally, the constant comoving case (dashed line) produces a more extended reionisation history, still it can only account for $\lsim 27\%$ of the ionised volume at $z\gsim 6$. In this case, reionisation by $z\gsim 6$ requires a mean escape fraction from star forming galaxies $\gsim 10\%$. 

It is interesting to note that the benchmark case (as well as the AGNs observed by \citet{giallongo14}) would produce a 
\HI ionisation rate consistent with $z\simeq 5-6$ data \citep[e.g., ][]{wyithe11,calverley11}, still it falls short in reionsing the IGM at 
$z\gsim 6$ (Fig.~\ref{fig2}). In other words, matching the observed level of the ionising background just below the ionisation 
redshift does not guarantee that a particular model is actually consistent with the entire reionisation history of the IGM. 

A final comments concerns \HeII~reionisation. Though a detailed assessment of such process is beyond the scope of this Letter, an AGN dominated background would certainly lead to an extended reionisation epoch. This may agree with the recent claiming of \citet{worseck14}, but it may be in conflict with the sharp increase of the IGM temperature at mean cosmic density observed in the range $2\lsim z \lsim 4$ \citep{schaye00,becker11,bolton12,bolton14,boera14}.  

\section{Conclusions}

Under reasonable assumptions, we have shown that a population of unobscured, UV emitting AGNs at $z\gsim 5$ if 
leading \HI reionisation would exceed observational constraints derived from the unresolved fraction of the X-ray background. Even a constant comoving emissivity at high-$z$ would not be enough to produce an AGN dominated ionising background. AGNs can account for a fraction of the ionising photon budget $\lsim 13\%$, calling for a dominant contribution from star forming galaxies. Given the observational constraints \citep[e.g.,][]{bouwens11} on the galaxy population at high-$z$, this in turns requires a large mean escape fraction ($\gsim 10\%$).

\acknowledgements
{We thank A. Comastri, P. Madau, A. Moretti for many fruitful discussions, and E. Giallongo for allowing us to use  
their results  before publication.}

\bibliographystyle{aa}
\bibliography{/Users/francescohaardt/Documents/Science/mybib}

\end{document}